\documentclass[seceq,supplement]{ptptex}

\usepackage{graphicx}

\title{
Replica Monte Carlo Simulation (Revisited)%
}

\author{
Jian-Sheng \textsc{Wang}$^{(1)}$ and Robert H. \textsc{Swendsen}$^{(2)}$
}

\inst{
$^{(1)}$Department of Computational Science,
National University of Singapore, Singapore 117543, 
Republic of Singapore
\\
$^{(2)}$Department of Physics, Carnegie Mellon University,
Pittsburgh, PA 15213, USA
}

\recdate{July 11, 2004}

\abst{
In 1986, Swendsen and Wang proposed a replica Monte Carlo algorithm
for spin glasses [Phys. Rev. Lett. {\bf 57} (1986) 2607].  Two
important ingredients are present, (1) the use of a collection of
systems (replicas) at different of temperatures, but with the same
random couplings, (2) defining and flipping clusters. Exchange of
information between the systems is facilitated by fixing the $\tau$
spin ($\tau=\sigma^1\sigma^2$) and flipping the two neighboring
systems simultaneously.  In this talk, we discuss this algorithm and
its relationship to replica exchange (also known as parallel
tempering) and Houdayer's cluster algorithm for spin glasses.  We
review some of the early results obtained using this algorithm.  We
also present new results for the correlation times of replica Monte
Carlo dynamics in two and three dimensions and compare them with
replica exchange.  }

\begin{document}

\maketitle

\section{Introduction}

The spin glass problem has been studied extensively over the past 30
years \cite{Binder-Young,Parisi,Kawashima-Rieger}.  Monte Carlo
simulation has been one of the main tools.  However, spin glass models
are among the most difficult to simulate, due to their extremely slow
dynamics at low temperatures.  Unlike the ferromagnetic models
\cite{SW}, cluster algorithms for spin glasses have only limited
success.

In 1986, a cluster algorithm by the name ``replica Monte Carlo'' was
proposed for spin glasses \cite{replicaMC}.  Although several other
algorithms were later developed
\cite{Liang,Houdayer,exchange,Marinari}, replica Monte Carlo
remains one of the best algorithm for spin glasses.  Replica Monte
Carlo works very well in two dimensions, reducing the correlation time
enormously comparing to single spin flips. In one dimension, it is
similar to the Swendsen-Wang cluster algorithm with a constant
correlation time independent of system size.  In three and higher
dimensions, it still provides a great improvement over single spin
flips, although improvement is not as dramatic. As we will show here,
replica Monte Carlo becomes essentially equivalent to ``replica
exchange'' or ``parallel tempering'' of Hukushima and Nemoto
\cite{exchange} in three or higher dimensions.

\section{Replica Monte Carlo}

Replica Monte Carlo actually contains two ideas, (1) the simultaneous
simulation of a collection of systems at different temperatures, (2)
cluster moves defined through the $\tau$ spin, where $\tau
=\sigma^1\sigma^2$ is the overlap of the systems at nearby
temperatures.  The first idea was later used in parallel tempering and
also in a similar proposal by Geryer \cite{Geryer}.

The reason for simulating an array of systems with the same random
couplings at different temperature $T_i$ is that configurations can be
rapidly equilibrated at high temperatures and the information can be
quickly transferred to low temperatures.  Exchanges are carried out
pairwise for systems at neighboring temperatures.  Consider two
replica configurations $\sigma^1$ and $\sigma^2$.  The joint
distribution of the two spin configurations $\sigma^1$ and $\sigma^2$
is a product of the two independent Boltzmann distributions at
temperature $T_1$ and $T_2$, respectively.  The combined system has
the ``Hamiltonian''
\begin{equation}
H_{{\rm pair}}(\sigma^1,\sigma^2) =  - \sum_{\langle i,j\rangle} \left( 
\beta^1 J_{ij} \sigma^1_i \sigma^1_j + \beta^2 J_{ij} \sigma^2_i \sigma^2_j \right),
\end{equation}
where $\beta^1=1/(kT_1)$ and $\beta^2=1/(kT_2)$, with a probability
distribution proportional to $\exp\bigl(-H_{{\rm pair}}(\sigma^1,
\sigma^2)\bigr)$.  Clearly, any Monte Carlo moves that keep the above
distribution invariant are valid.  Specifically, we can design Monte
Carlo moves that satisfy detailed balance with respect to the joint
distribution.  There are many possible such choices.

In our original replica Monte Carlo implementation
\cite{replicaMC,replicaMCPRB,replicaMCreview}, we proposed 
cluster moves.  The clusters are defined by connected region of the
same $\tau$-spin, $\tau_i = \sigma_i^1 \sigma_i^2$.  Two nearest
neighbor sites $i$ and $j$ are considered connected and in the same
cluster if $\tau_i = \tau_j$.  To keep the definition of the
$\tau$-cluster invariant with respect to the move, we must
simultaneously flip both systems.  More precisely, we rewrite the pair
Hamiltonian in the form
\begin{equation}
H_{{\rm pair}}(\sigma^1;\tau) =  - \sum_{\langle i,j\rangle} ( 
\beta^1 + \beta^2 \tau_i \tau_j ) J_{ij} \sigma^1_i \sigma^1_j.
\end{equation}
Since $\tau$-spins are fixed, the system is a new spin glass with
interaction strength proportional to $\beta^1 + \beta^2$ inside a
$\tau$-cluster and decreased to $\beta^1 - \beta^2$ between the
$\tau$-clusters.  In particular, if $\beta^1 = \beta^2$ these $\tau$
clusters are free to flip without costing any energy.  In such a case,
the $\tau$ clusters also have a good physical meaning -- they
correspond to the Edwards-Anderson order parameter for spin glass.
The interactions between the clusters are governed by the above
Hamiltonian and can be rewritten as an effective Hamiltonian between
clusters.  Let $\eta_a = 1$ be the initial spin value for all cluster
$a$ indicating non-flipping, and $\eta_a=-1$ indicating a flip of the
cluster, the effective interaction Hamiltonian between clusters is
\begin{equation}
H_{{\rm cls}}(\eta) = - \sum_{a,b} k_{a,b} \eta_a \eta_b,
\end{equation}
where the effective coupling is from the sum of the couplings along
the boundary between cluster $a$ and $b$,
\begin{equation}
k_{a,b} = ( \beta^1 - \beta^2  ) \!\!\!
\sum_{\langle i \in a, j \in b\rangle} \!\!\!
J_{ij} \sigma^1_i \sigma^1_j.
\end{equation}
Note that the $\sigma$ above should take the value before the cluster
flips.  We can now simulate the above cluster model with any valid
Monte Carlo algorithm, e.g., Metropolis single spin flip, or even
Swendsen-Wang cluster flip.  After the cluster update to variables
$\eta_a$, the original spins are changed according to $\sigma^1_i
\leftarrow \eta_a \sigma^1_i$ and $\sigma^2_i \leftarrow \eta_a
\sigma^2_i$ for $i \in a$.  The replica cluster moves must be
supplemented by a normal Metropolis simulation of $\sigma^1$ and
$\sigma^2$ to break the conservation of $\tau$.  The Metropolis step
makes the algorithm ergodic.

In our earlier implementation, we made a sequential sweep of all the
clusters with a heat-bath rate.  It is interesting to note that the
replica exchange (parallel tempering) is equivalent to a trial move
where all the $\tau = -1$ clusters are flipped, or a move where all
$\tau = +1$ clusters are flipped, followed by a global sign flip.
Since the $\tau$ spins are constant with respect to the replica Monte
Carlo moves, any move based on values of $\tau$ satisfies detailed
balance and is thus valid.  Iba gave further discussion for the
relation between replica Monte Carlo and other extended ensemble
methods \cite{Iba}.  Another connection is with the cluster algorithm
of Houdayer \cite{Houdayer}.  In Houdayer's implementation
\cite{houdayer-imp}, a site is picked at random, the associated
$\tau$-cluster with that site is formed and flipped with probability 1
for systems with same temperature.  This is similar to the Wolff
variation \cite{Wolff} of the Swendsen-Wang algorithm.  For systems
between different temperatures, replica exchange was used.

Another improvement that was suggested in 1986, but not implemented
\cite{replicaMCPRB}, is 
to introduce a Swendsen-Wang step within the $\tau$ clusters.  In such
case, each bond has interaction strength $( \beta^1 + \beta^2 ) J_{ij}
\sigma^1_i \sigma^1_j$.  The presence of a bond is set with
probability $P = 1 - \exp\bigl(-2(\beta^1+\beta^2)|J_{ij}|\bigr)$ if
the interaction is satisfied, i.e., $J_{ij} \sigma^1_i \sigma^1_j>0$.
The bond is absent otherwise.  We do not apply this step between the
$\tau$ clusters.  This further breaks up the $\tau$ clusters.  One
technical advantage of having this step is that ergodicity can be
maintained without evoking a single-spin-flip sweep.  Although it is a
great help at high temperatures, at interesting temperatures for spin
glass phase transitions, $\beta \approx 1$, the value of $P \approx
0.98$ is too large.  The dynamics is nearly the same as the original
version.

As it is obvious, most of the CPU time is devoted to the computation
of the effective coupling, $k_{a,b}$.  This can be done efficiently at
about $O(N)$ where $N$ is the number of spins of the system.  This is
because that the number of possible interactions is bounded by $Nd$
where $d$ is the dimension of the system. We can pre-allocate
sufficient memory based on cluster size and the total number of
clusters.  In a new implementation of the algorithm, on a 1GHz Itanium
2, each replica Monte Carlo sweep (MCS) per spin takes 1.06 $\mu$sec
(on a $128\times128$ lattice).  This should be compared with a
straightforward Metropolis algorithm, which runs at 0.29 $\mu$sec per
MCS per spin.  Little size dependence is seen on the speed.

\section{Some Early Results}

The replica Monte Carlo was used to compute two-dimensional spin-glass
susceptibility \cite{replicaMC} and low-temperature heat capacity
\cite{replicaMCPRB} for the $\pm J$ spin glass model, where the
coupling $J_{ij}$ takes $+J$ or $-J$ with equal probability. The spin-glass
susceptibility was analyzed in the form of power-law divergence $\chi
\sim T^{-\gamma}$, assuming $T_c=0$.  The exponent $\gamma \approx
5.1$ was found.  Recent work \cite{Houdayer} seems to support $\chi
\sim \exp(2(2-\eta)\beta J)$ instead, where $\eta$ is the critical 
exponent associated with correlation function, $g(r) \sim r^{-\eta}$.

The low-temperature heat capacity is more interesting.  We found
\cite{replicaMCPRB} that $c \sim \beta^{2} \exp(-2\beta J)$ from our
simulations and argued that the elementary excitations in $\pm J$
Ising spin glass are kink/antikink pairs.  This result was disputed
\cite{Kardar}, but recent more extensive calculations with exact
algorithms for partition function \cite{exact-partition} supported our
result.

We also used the replica Monte Carlo algorithm with a renormalization
group analysis \cite{replicaMCRG}.  The results show an early
confirmation of the absence of a phase transition at finite
temperature in two dimensions, and clear evidence of a
finite-temperature spin-glass phase transitions in three and four
dimensions.

\begin{figure}
\includegraphics[width=0.8\columnwidth]{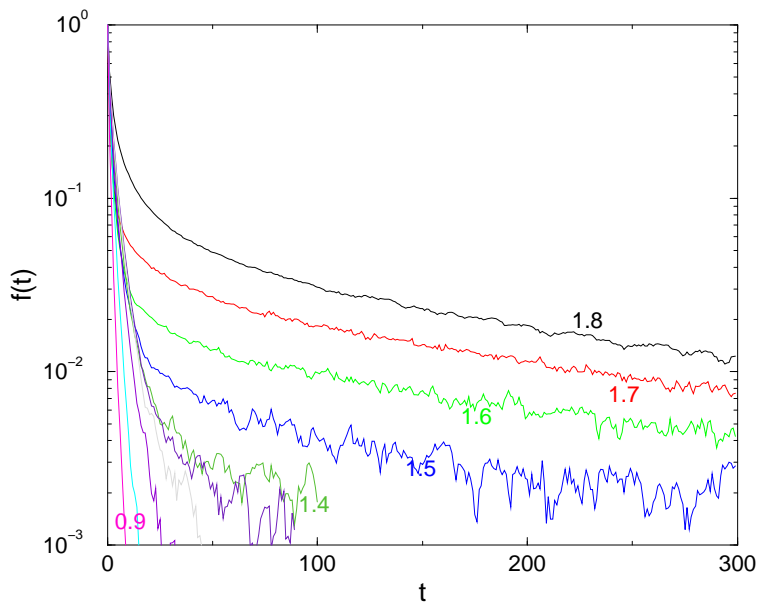}
\caption{\label{fig1} Time correlation function for the order parameter
$q$ on a $128 \times 128$ $\pm J$ Ising spin glass system, using
replica Monte Carlo with $\beta J$ distributed from 0.1 to 1.8 in
steps of 0.1 (only 0.9 to 1.8 are plotted).  }
\end{figure}

\begin{figure}
\includegraphics[width=0.8\columnwidth]{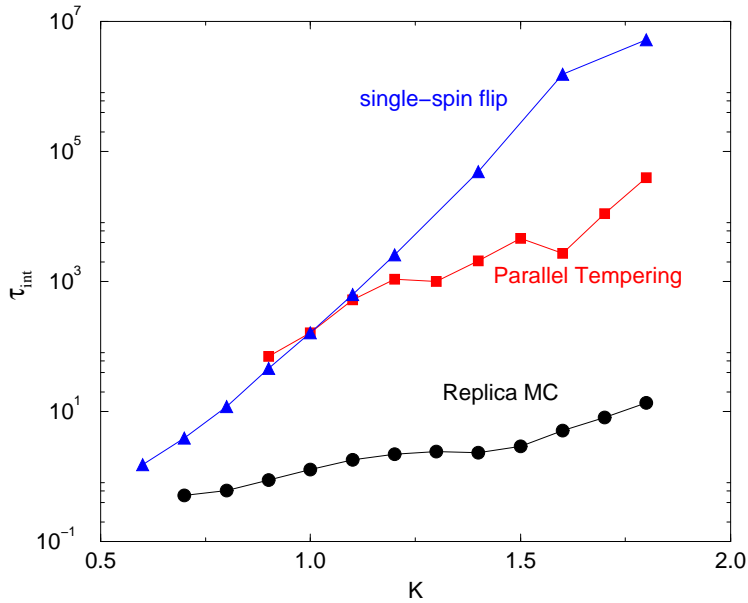}
\caption{\label{fig2} Comparison of integrated correlation time
on a $128 \times 128$ lattice for single-spin-flip (triangles),
parallel tempering (squares), and replica Monte Carlo (circles).  The
$K=\beta J$ value is distributed from 0.1 to 1.8 in spacing of 0.1.
Typically, $10^6$ MCS are used.  }
\end{figure}

\begin{figure}
\includegraphics[width=0.8\columnwidth]{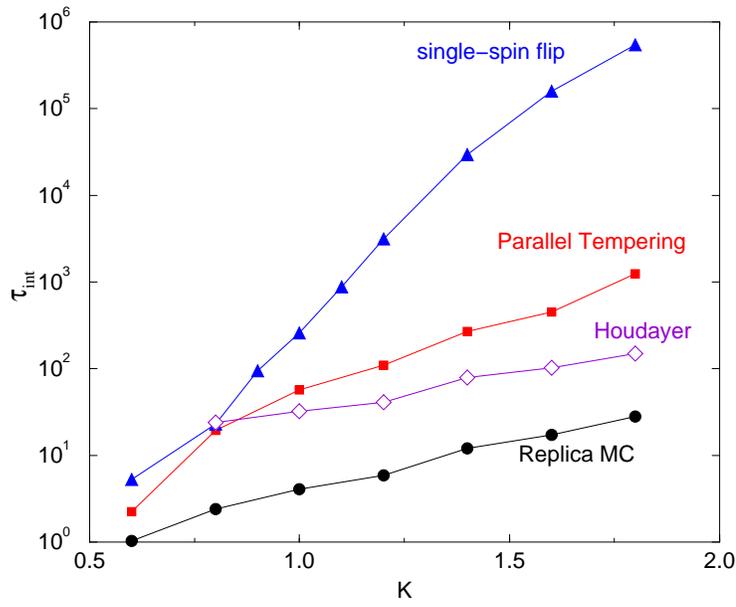}
\caption{\label{fig3} Comparison of integrated correlation times
on a $32 \times 32$ lattice for several algorithms.  The $K=\beta J$
value is distributed from 0.2 to 1.8 in spacing of 0.2.  }
\end{figure}

\begin{figure}
\includegraphics[width=0.8\columnwidth]{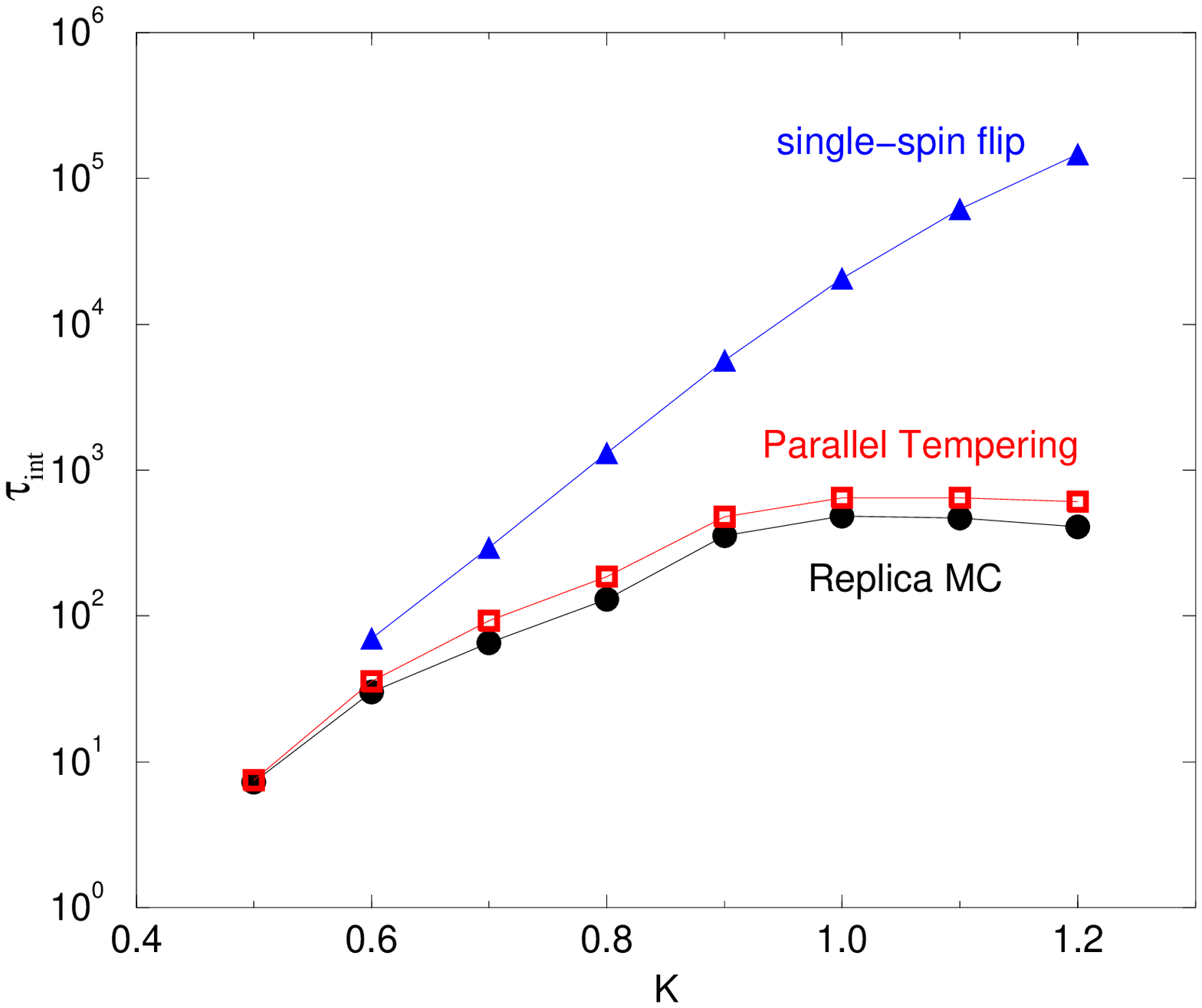}
\caption{\label{fig4} Integrated correlation time
on a $12 \times 12 \times 12$ lattice for single-spin-flip, parallel
tempering, and replica Monte Carlo.  The $\beta J$ value is
distributed from 0.1 to 1.2 in spacing 0.1.  }
\end{figure}

\section{Relaxation Times and Comparison}
In order to compute the spin-glass order parameter and susceptibility,
two sets of replica are used.  In each sweep, we do one
single-spin-flip sweep for the two sets, and replica Monte Carlo
between nearby temperatures for both sets, and replica Monte Carlo at
same temperature between the sets.  Swendsen-Wang clusters are
generated within a $\tau$-cluster.  The Edwards-Anderson order
parameter appears to contain the slowest relaxation mode (the energy
correlations decay faster).  Thus we compute (for a single realization
of random coupling)
\begin{equation}
f(t) = { \langle q(t) q(0) \rangle - \langle q(0) \rangle^2 \over
        \langle q(0)^2 \rangle - \langle q(0) \rangle^2 },
\end{equation}
where $q = (1/N) |\sum_i \tau_i|$, $\tau_i$ is the overlap spin at the
same temperature.

In figure \ref{fig1}, we present correlation functions at some typical
temperatures.  The function has an initial fast decay followed by a
slow relaxation at long time.  This causes a factor of 10 difference
between the integrated correlation time, defined by
\begin{equation}
 \tau_{{\rm int}} = \int_0^\infty \!\! f(t)\, dt,
\end{equation}
and exponential correlation time, $f(t) \approx A \exp(-t/\tau)$, $t
\to \infty$.  The initial fast decay reflects the quick swap of
configurations between neighboring temperatures.  However, they can be
swapped back again, leading to a longer exponential correlation times.
In any case, the integrated correlation time is relevant to the error
in sample average.  In figure \ref{fig2}, we present $\tau_{{\rm
int}}$ for one single random sample of $128 \times 128$.  In contrast
to single-spin-flip and parallel tempering, remarkably small
correlation times are observed for the replica Monte Carlo algorithm.

In figure \ref{fig3}, we compare several algorithms at $32 \times 32$.
Comparing to the results in fig.~\ref{fig2}, we found that size effect
is small.  The curve labelled Houdayer is from a simulation of 2 sets
of systems (instead of a large number of sets).  A unit time step
consists of one Metropolis sweep for all the systems for the two sets,
single cluster moves between the sets, and replica exchanges in a set
between neighboring temperatures.

In figure \ref{fig4}, we present the correlation times for a
three-dimensional $\pm J$ Ising spin-glass system of size $12^3$.
The replica Monte Carlo includes moves between the two sets at same
temperature.  As one can see, the performance of replica Monte Carlo
and parallel tempering is nearly the same.  This is because in three
or higher dimensions, there are essentially only two $\tau$ clusters,
one with $+$ $\tau$-spin and other $-$ $\tau$-spin.  Flipping either
of them is equivalent to exchange of configurations.

In these correlation time comparisons, we have not fine-tuned the
simulation parameters (distribution of temperatures, number of sweeps
for each type of move, and number of sets).  Future research on the
optimal simulation parameters might reveal further improvements in
efficiency.

\section{Conclusion}
We introduced the replica Monte Carlo algorithm and presented new
correlation time data.  These results show that replica Monte Carlo
algorithm is extremely efficient in two dimensions, and is even much
better than parallel tempering.  In three dimensions, replica Monte
Carlo and parallel tempering are of comparable efficiency.


%


\begin{thebibliography}{99}
\bibitem{houdayer-imp} Houdayer used a two-dimensional array of
systems -- $M$ systems at each temperature.  Single cluster moves were
performed for systems only at the same temperature, and replica
exchange was used between different temperatures.  We argue that a
replica Monte Carlo move would be more efficient for systems with
different temperatures.
\end{thebibliography}
\end{document}